# EscalNet: Learn isotropic representation space for biomolecular dynamics based on effective energy

Guanghong Zuo[1*]

[1] Wenzhou Institute, University of Chinese Academy of Sciences,
Wenzhou, Zhejiang 325001 China
[*] Corresponding author E-mail: ghzuo@ucas.ac.cn (Zuo GH).
ORCID: 0000-0002-7822-5969.

**ABSTRACT**：Deep learning has emerged as a powerful framework for analyzing biomolecular dynamics trajectories, enabling efficient representations that capture essential system dynamics and facilitate mechanistic studies. We propose a neural network architecture incorporating Fourier Transform analysis to process trajectory data, achieving dual objectives: eliminating high-frequency noise while preserving biologically critical slow conformational dynamics, and establishing an isotropic representation space through the last hidden layer for enhanced dynamical quantification. Comparative protein simulations demonstrate our approach generates more uniform feature distributions than linear regression methods, evidenced by smoother state similarity matrices and clearer classification boundaries. Moreover, by using saliency score, we identified key structural determinants linked to effective energy landscapes governing system dynamics. We believe that the fusion of neural network features with physical order parameters creates a robust analytical framework for advancing biomolecular trajectory analysis.

# 1. Introduction

Molecular dynamics simulations offer atomic-scale resolution of biomolecular motions, permitting researchers to systematically investigate protein dynamical mechanisms with unprecedented detail [1–11]. These computational techniques have emerged as an indispensable tool in contemporary protein dynamics research [12–26]. An expanding body of scholarly work focuses on identifying functionally relevant collective motion patterns, particularly slow modes of conformational changes, from extensive molecular dynamics trajectory datasets to establish physically meaningful dynamical representations [27–31]. The classical theoretical framework for protein conformational transitions, grounded in equilibrium statistical mechanics, theorizes protein folding processes as stochastic walks across multidimensional funnel-shaped free energy landscapes characterized by rugged topography [32]. Conversely, kinetic network models utilize stochastic dynamics formalism to characterize conformational transitions as Markovian jumps between discrete metastable states within a network topology [33–37]. This methodological paradigm addresses the high-dimensionality constraints intrinsic to free energy landscape visualizations while quantitatively describing non-equilibrium dynamical phenomena [38].

The integration of deep learning methodologies with molecular dynamics trajectory analysis capitalizes on the inherent spatial mapping capacities of deep neural networks to construct feature spaces that effectively capture slow dynamical modes inherent in complex systems [39–42]. Subsequent statistical interrogation of these learned representations enables comprehensive interpretation of system dynamics through free energy landscapes and kinetic network analyses. This methodological paradigm facilitates iterative conformation sampling within the feature space and enables strategic implementation of bias potentials in the learned representation space, thereby achieving enhanced sampling efficiency [43]. Schneider et al. demonstrated that artificial neural networks trained on enhanced sampling datasets enable compact representation and precise computation of high-dimensional free energy surfaces, permitting accurate prediction of high-dimensional thermodynamic observables in complex molecular systems [44]. Ketkaew et al. established an unsupervised deep autoencoder neural network (DAENN) architecture capable of extracting transferable reaction coordinates from trajectory data, enabling autonomous exploration of hidden metastable states and free energy surfaces through metadynamics simulations in machine-learned latent spaces [45]. Noé and coworkers proposed the Boltzmann generator [46], utilizing Real NVP normalizing flow architectures, achieves efficient sampling of rare critical configurations through the integration of flow-based generative models with replica exchange methodologies, leading to the development of Learning-Replica Exchange (LREX) enhanced sampling protocols [47]. Gao and coworkers pioneered the Information Distillation of Metastability (IDM) approach, which combines dimensionality reduction and clustering techniques in deep unsupervised learning frameworks to identify metastable states during phase transitions [48]. Pande and coworkers designed variational dynamics encoder (VDE) architecture, constructed using variational autoencoder principles, incorporates saliency map analysis to pinpoint functionally critical residues in biomolecular conformational dynamics [49].

Most of existing methodological frameworks predominantly focus on deep learning architectures demonstrate insufficient attention to symmetry of representation space, particularly regarding uniformity within individual dimensions and metric consistency across different dimensional axes. Notably, for physical systems exhibiting specific characteristics, the integration of physically meaningful order parameters through feature engineering not only imbues the analytical space with thermodynamic interpretability but also optimizes its symmetry properties, thereby significantly improving machine learning algorithm efficiency. In our previous works, we formulated the Effective Energy Rescale Space Trajectory Mapping (EspcTM) method, which achieves systematic integration of Fast Fourier Transform (FFT) and multivariate linear regression algorithms [50]. This framework enables extraction of low-frequency effective energy – a crucial order parameter characterizing slow dynamic processes in complex systems – from total energy profiles. Through dimensional rescaling of original system features using this low-frequency component, we successfully established metrical uniformity across dimensions in the representation space. Application of EspcTM has permitted successful construction of conformational transition networks for both Brownian particle systems and alanine-12 peptide ensembles. However, while the linear regression framework in EspcTM ensures metric consistency between dimensions, it remains inadequate in addressing uniformity requirements within individual dimension. The attainment of such dimensional uniformity necessitates implementation of nonlinear scaling paradigms, where neural network architectures – with their inherent nonlinear representational architecture and relatively straightforward trainability – present a theoretically viable computational framework for achieving this objective.

In this study, we propose a deep neural network approach to characterize the nonlinear relationship between conformational features of biomolecules and effective energy of molecular dynamics trajectories, and constructed a representation space scaled by the effective energy metrics. This framework is applied to analyze molecular dynamics trajectories of the fast-folding protein domain Chignolin [51,52]. Our results demonstrate that compared with linear regression methods, the neural network-based approach generates an isotropic representation space, thereby yielding smoother state similarity matrices and enabling more distinct classification of system states. Furthermore, to enhance the interpretability of the network, we employ saliency score to quantify the contribution of individual features to molecular dynamical processes, successfully identifying key determinants of conformational transitions. We believe that the integration of neural networks with physical order parameter analysis offers a novel paradigm for molecular trajectory investigation, which may significantly advance the mechanistic understanding of biomolecules.

## 2. Models and Methods

### 2.1 Molecular Dynamics

We elucidate the analytical methodology using the dynamic behavior of the commonly studied fast-folding protein domain Chignolin (PDB: 5AWL) as a representative example [51]. The structure of Chignolin was retrieved from the Protein Data Bank and solvated in a dodecahedron periodic box, maintaining a minimum distance of 10 Å between the solute and periodic boundary. The system incorporated 1926 water molecules and two sodium ions to neutralize the protein's net charge, resulting in a total of 5946 particles modeled using the AMBER99SB-ILDN force field [53]. Molecular dynamics simulations were performed using the GROMACS software package [54]. During simulations, covalent bonds involving hydrogen atoms were constrained via the LINCS algorithm with a 2.0 fs time step. Non-bonded interactions were truncated at 15.0 Å, while long-range electrostatic interactions were treated using the Particle Mesh Ewald (PME) method [55]. Prior to production runs, the system underwent energy minimization followed by a 100 ns equilibration in the NPT ensemble to achieve appropriate density for subsequent NVT ensemble simulations. For trajectory analysis in this study, we employed a 1.0 μs molecular dynamics trajectory simulated at 375 K [52].

**2.2 Feature Extraction**

A 1.0 μs molecular dynamics (MD) simulation trajectory was systematically sampled at 10.0 ps intervals, resulting in $N_t = 100,000$ frames. To characterize state transitions of the system, the trajectory was projected into a space spanned by $N_b$ basis functions $\{\hat{A}^\mu(\vec{q})\}_{\mu=1,...,N_b}$. Consequently, each trajectory was represented as an $N_t \times N_b$-dimensional feature matrix in the Hilbert space, formally expressed as:

$$V = \left(\hat{A}^1(\vec{q}), \hat{A}^2(\vec{q}), \hat{A}^3(\vec{q}), ..., \hat{A}^{N_b}(\vec{q})\right)$$

Where $\vec{q}$ denotes structural metrics. For protein conformational transitions, torsional angles $(\varphi, \psi)$ of the peptide backbone or inter-residue distances between $C_\alpha$ atoms represent appropriate choices. In this study, we employed backbone torsion angles $(\varphi, \psi)$ as the fundamental collective coordinates. To effectively mitigate periodic boundary artifacts inherent in angular measurements, the basis functions $\{\hat{A}^\mu(\vec{q})\}_{\mu=1,...,N_b}$ were constructed using sine and cosine transformations of the torsional angles [56]. The Chignolin protein domain contains 8 pairs of backbone torsion angles, yielding $N_b = 32$ basis functions following this transformation protocol.

**2.3 Low Frequency Effective Energy**

To capture the slow motion of the system, we introduce a low-frequency effective energy as an order parameter for external scaling to characterize system evolution. The specific protocol involves performing a Fast Fourier Transform (FFT) on the total energy of the system to obtain frequency-domain spatial vectors [57]. Retaining the first K frequency components followed by inverse transformation yields the low-frequency energy:

$$\widetilde{\omega}_k = \sum_{n=0}^{N_t-1} \varepsilon_n \cdot e^{-in\omega_k} \quad \Rightarrow \quad n_K^{\tilde{\varepsilon}} = \sum_{k=0}^{K-1} \widetilde{\omega}_k \cdot e^{in\omega_k}$$

Where $K \ll N_t$. If the characteristics of the system are well-defined, $K$ value can be provided as a hyperparameter. In this study, we introduce a data-centric methodology to automate the process of hyperparameter tuning, thereby eliminating the need for manual input. Molecular dynamics simulations reveal that solvation effects are crucial in determining the slow motion modes of biomolecules. We therefore employ the biomolecule's collective coordinates $V = \left(\widehat{A}^1(\vec{q}), \widehat{A}^2(\vec{q}), \widehat{A}^3(\vec{q}), \ldots, \widehat{A}^{N_b}(\vec{q})\right)$ to infer low-frequency energy components:

$$\tilde{\varepsilon}^K = f(V) + \epsilon^K$$

Through iterative fitting across multiple K values, we optimize the model parameters and select the optimal K value maximizing the coefficient of determination:

$$K^* = arg\,max \sqrt{1 - \frac{\overline{(\epsilon^K)^2}}{(\sigma^K)^2}}$$

where $\epsilon^K$ denotes the fitting residual and $(\sigma^K)^2$ represents the variance of system energy at frequency K. This methodology enables automated identification of dynamically relevant low-frequency modes while maintaining physical interpretability.

### 2.4 Neural Network

During the fitting process, the selection of an appropriate function $f(V)$ significantly influences the efficiency of representation. In our preliminary studies, we used linear functions as fitting functions. However, a nonlinear relationship evidently exists between collective coordinates and low-frequency energy components. In this work, we employ a neural network approach to capture this nonlinear relationship. As illustrated in Figure 1, the network architecture comprises three fully connected layers. The input layer dimension is data-determined, i.e. $n = N_b = 32$. The node numbers in the hidden layer $m$ and feature layer $s$ are system-dependent, with both parameters set to match the system's degrees of freedom $m = s = 16$ in this investigation. In the neural network architecture, all layers except for the feature-to-output connection employ ELU activation functions, which are known for their ability to prevent the dying ReLU problem and have been shown to enhance training efficiency and accuracy. The model is trained using the Adam optimizer, which is known for its adaptive learning rate capabilities. The initial learning rate is set to 0.001, a batch size of 10 is used for processing, and the mean squared error (MSE) function is employed as the loss criterion. Ridge regression is implemented through weight decay regularization. The entire neural network framework is constructed using PyTorch.

Upon obtaining the optimized model and optimal cutoff frequency $K^*$ through training. K-Means, a widely-used clustering algorithm, was applied to the vectors of the last hidden layer to segment the dynamical process into distinct states. Notably, the absence of activation functions in the output layer ensures that the effective energy equals the summation of last hidden layer, maintaining consistent scaling between feature magnitudes and energy values.

## 3. Results and Discussion

Figure 2A presents the root mean square deviation (RMSD) profile of Chignolin relative to its native conformation obtained from molecular dynamics simulations conducted at 375 K. The trajectory exhibited two prominent transitional events along with multiple short-lived conformational fluctuations. This observation highlights the protein's gradual transition away from its native state, traversing distinct conformational substates during the simulation. The recorded multi-state dynamical behavior constitutes a robust dataset for implementing advanced trajectory clustering methodologies in subsequent analyses.

### 3.1 Low-Frequency Effective Energy

By applying low-frequency cutoff filtering, we obtained the summed energy of all low-frequency bands across specified cutoff frequencies (see Methods for technical details). Using these low-frequency energy values as the target function for neural network training, we established optimal fitting relationships between low-frequency energy at different cutoff frequencies and characteristic parameters, while calculating corresponding multiple correlation coefficients for each frequency (see Figure 2C). For comparative analysis, the frequency-dependent multiple correlation coefficients from linear regression are also presented in Figure 2C. Both methods exhibited similar overall trends, but the nonlinear components of the neural network significantly reduced fitting errors. This indicates that linear terms primarily mediate global modulation of dimensional metric consistency, which dominates the fitting process and explains the analogous trend patterns between methods. However, the nonlinear effects introduced by the neural network implement localized fine-tuning through uniformity modulation within individual dimensions, thereby complementing the fitting performance. Consequently, the neural network achieved systematically higher correlation coefficients across all frequencies, demonstrating the essential contribution of nonlinear terms.

Based on the multiple correlation coefficients shown in Figure 2C, we selected the neural network model trained at the optimal cutoff frequency (11.5 MHz). The energy values output by this optimized model were defined as effective energy. Figure 2B illustrates the temporal evolution of effective energy (red curve) compared with total system energy (cyan curve). Notably, these two energy profiles exhibit fundamental differences: while total system energy fluctuates extensively across a broad range – theoretically containing conformational change information but obscured by stochastic noise – the effective energy derived from neural network analysis displays discrete transitions between distinct energy states, effectively filtering out noise to reveal systematic patterns. Further comparison with the RMSD trajectory in Figure 2A demonstrates analogous temporal patterns between effective energy and structural deviations. This confirms that effective energy captures essential features of protein conformational transitions while providing enhanced state discrimination compared to RMSD. For instance, the segments at 0.4–0.5 μs and 0.6–0.8 μs exhibit comparable RMSD averages that differ only in fluctuation magnitude, whereas their effective energy values show distinct plateau levels, enabling direct state differentiation. These results establish effective energy as a robust order parameter for characterizing protein conformational transitions.

**3.2 Analysis of Kinetic Processes**

The last hidden layer of the neural network provides an optimized representational space for protein conformations (red nodes in Figure 1). In this architecture, the effective energy output (purple node) equals the summation across all dimensions of the representational space (red nodes). Consequently, the gradient of the effective energy with respect to any dimension remains constant at unity, indicating that after sufficient training, all dimensions share identical scaling and exhibit intrinsic uniformity. Owing to the high optimization of this representation space, the simplest K-Means clustering algorithm can effectively partition the entire kinetic trajectory [58]. Figure 3A illustrates the kinetic transitions among three distinct states. Protein folding occurs via intermediate state S2 during transitions between the folded state (S3) and the unfolded state (S1). The state transition pathway across the simulation trajectory (Figure 3B, upper panel) exhibits remarkable consistency with the patterns of RMSD and effective energy shown in Figures 2A and 2B.

The similarity matrices based on the representation by the last hidden layer were shown in Figure 3B. For comparison, we analyzed similarity matrices derived from two alternative approaches: a linearly regressed coordinate-aligned space (Figure 3C)

[50] and a standardized coordinate space without alignment (Figure 3D) [35]. While all matrices display comparable block partitioning, the raw coordinate matrix (Figure 3D) exhibits reduced discriminative power, as evidenced by predominant mid-spectrum coloration, leading to instability in clustering outcomes. The regression-aligned space (Figure 3C) improves discriminability and clustering robustness but it retains intra-block textural artifacts, causing artificial state-transition noise and trajectory fragmentation. In contrast, the neural network-derived representation space resolves these limitations comprehensively. It generates similarity matrices with enhanced discriminability while intrinsically eliminating intra-block textural noise, thereby enabling stable clustering without trajectory fragmentation. Notably, although trajectory smoothing algorithms can mitigate fragmentation artifacts, they introduce subjective hyperparameters (e.g., smoothing window size). The neural network approach provides a fully data-driven solution conducive to automated analysis.

Mechanistically, this superiority arises from differential contributions of reaction coordinates to conformational transitions. Independent standardization of coordinates assumes equal contribution from each dimension to conformational dynamics, rendering classification dependent on multidimensional kinetic cooperativity—a suboptimal strategy for stability. Global scaling via linear regression assigns coordinate-specific weights that reflect their relative contributions, improving discriminability (manifested as edge-biased coloration in similarity matrices). However, such global weighting fails to address intra-dimensional uniformity. The nonlinear transformations in the neural network enforce gradient uniformity (constant gradient of 1 to the effective energy) across all dimensions of the representation space. These properties eliminate trajectory fragmentation artifacts, produces smoother similarity matrices, and yields reasonable state partitioning.

### 3.3 Saliency Score Analysis

The preceding analysis demonstrates that the representation space derived from the last hidden layer of the neural network employs uniform scaling across and within its dimensions, thereby establishing an isotropic analytical space. This symmetry optimizes the analytical workflow, yielding more stable and reliable results. However, the nonlinear mapping between these dimensions and the original coordinates introduces challenges in interpretability. To address the limited interpretability of the neural network, we adopt the conceptual framework of Vanilla Gradient methods used in constructing saliency maps [59]. Specifically, we quantify the contribution of individual reaction coordinates to dynamical processes by statistically analyzing the gradients of the effective energy with respect to these coordinates. This approach aligns methodologically with the strategy employed by Pande and coworkers [49].

In practice, as the neural network is trained using backpropagation-based derivative computation, the gradients of the loss function with respect to the input variables can be easily obtained. Similar to how pixels with larger gradients of the loss function dominate image recognition in Vanilla Gradient methods, we use boxplot visualization to statistically describe the gradients of effective energy across all reaction coordinates (dihedral angles), as shown in Figure 4. The analysis reveals that $\psi 6$ and $\psi 8$ exhibit relatively greater contributions. This finding correlates with the dynamical modes identified earlier (Figure 3A): the studied trajectory features three distinct kinetic states corresponding to the zipping model of β-hairpin folding [51,60]. The transition from the unfolded to intermediate state is primarily governed by rotations of $\psi 6$, whereas the subsequent transition to the folded state is dominated by rotations of $\psi 8$.

## 4. Conclusion

In this study, we present EscalNet, a novel methodology for molecular dynamics trajectory analysis that integrates Fourier transform with neural networks. This approach enables the extraction of an energy-dimensional order parameter, termed effective energy, from the total energy output of molecular dynamics simulations, which effectively captures protein conformational changes. By interrogating the last hidden layer of the neural network, we establish a representation space optimized for analyzing protein conformational transitions. Through deliberate architectural design, we rigorously demonstrate that this representation space exhibits isotropic symmetry, a critical property underlying its exceptional capability to characterize system dynamics. We believe that leveraging physically meaningful order parameters to scale representation spaces represents an innovative paradigm with significant potential for enhancing deep representation learning, as evidenced by the growing body of research that explores the intersection of physics and deep learning. Furthermore, we develop a gradient-based metric quantifying the contribution of reaction coordinates to system dynamics through statistical analysis of effective energy profiles, thereby enhancing the interpretability of neural network models. The EscalNet methodology establishes a new conceptual framework for molecular trajectory analysis, offering substantial promise for advancing dynamical investigations of complex biological systems.

## Code availability

Demo code of EscalNet is available at https://gitee.com/ghzuo/escalnet

## Competing interests

The authors have declared that no competing interests exist.

## Acknowledgments

GHZ thanks the Wenzhou institute, University of Chinese Academy of Sciences (Grant No. WIUCASQD2021042).

# Figures Legends

**Figure 1 Schematic diagram of EscalNet.**

The original data were inputted from the orange node and passed through a fully connected network to the hidden layers (cyan and red nodes), and fitted the effective energy of the molecular system (purple node). The last hidden layer (red nodes) serves as the representation space for data analysis.

**Figure 2 Effective Energy Analysis**

(A) RMSD and (B) Effective energy (with cutoff of 11.5 MHz, red) and total energy (scaling down to 1/20 of the original for plotting, cyan) as a function of time for a typical simulation trajectory. (C) The multiple correlation coefficient as a function of cutoff of frequencies for neural network (EscalNet) and linear regression (EspcTM).

**Figure 3 State Transition and Similarity matrices of a typical Trajectory.**

(A) Three states transition. And similarity matrices obtained from feature space (B) rescaled by neural network (EscalNet), (C) rescaled by linear regression (EspcTM), and (D) without rescaling.

**Figure 4 Saliency score of the neural network.**

The boxplot of the derivatives of effective energy on original reactive coordinates (dihedral angles of the backbone). The orange horizontal line represents the median, the upper and lower edges of the blue model are the quartiles, the horizontal lines above and below are the boundaries, and the yellow dots are the mean values.

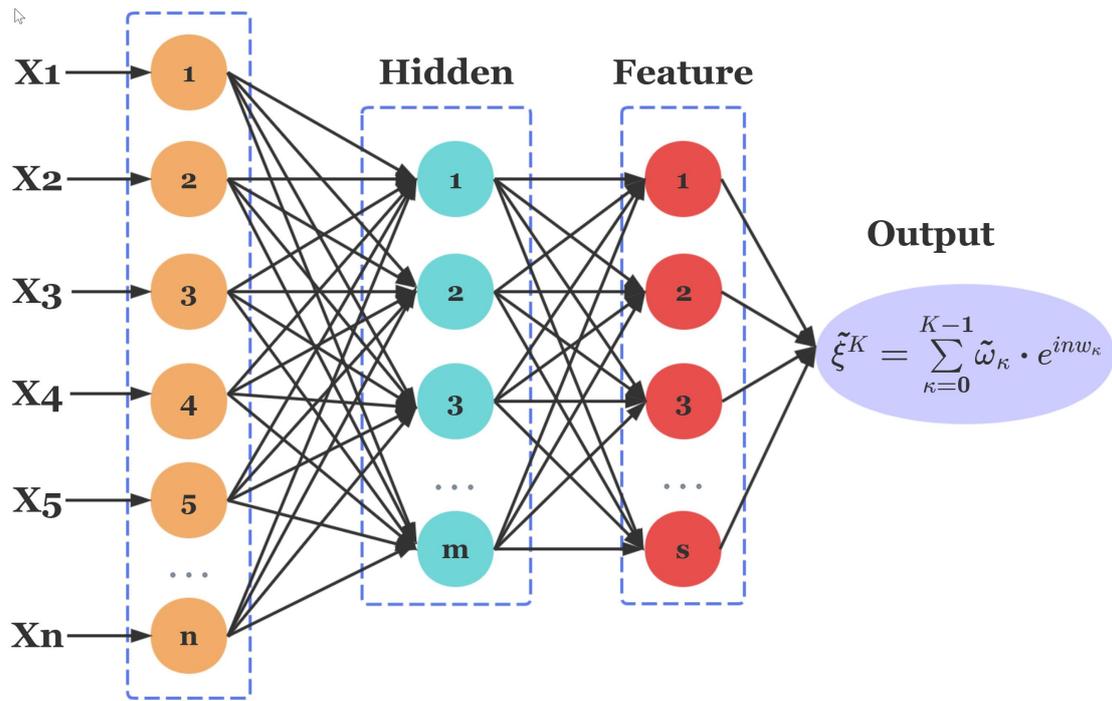

Figure 1 Schematic diagram of EscalNet.

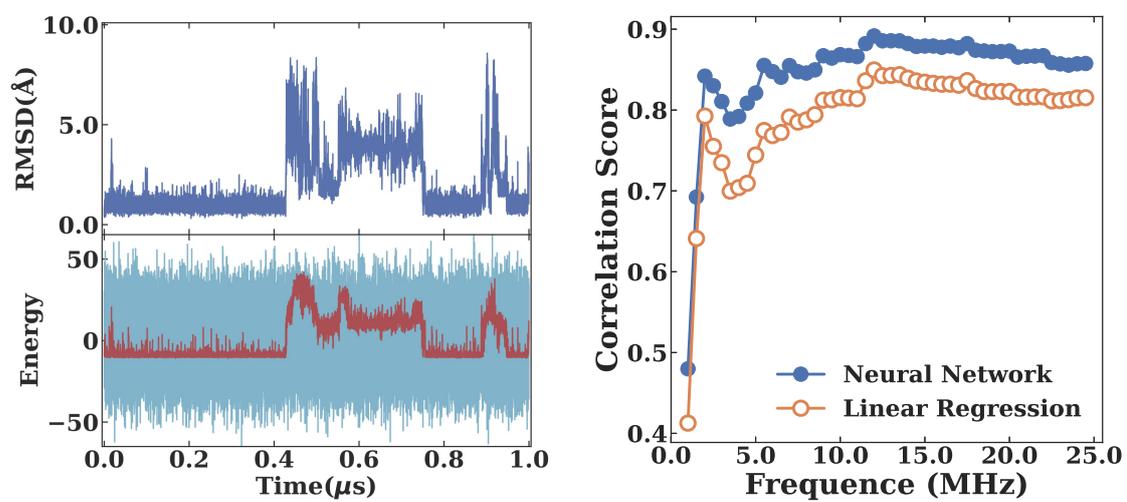

**Figure 2 Low Frequency Effective Energy**

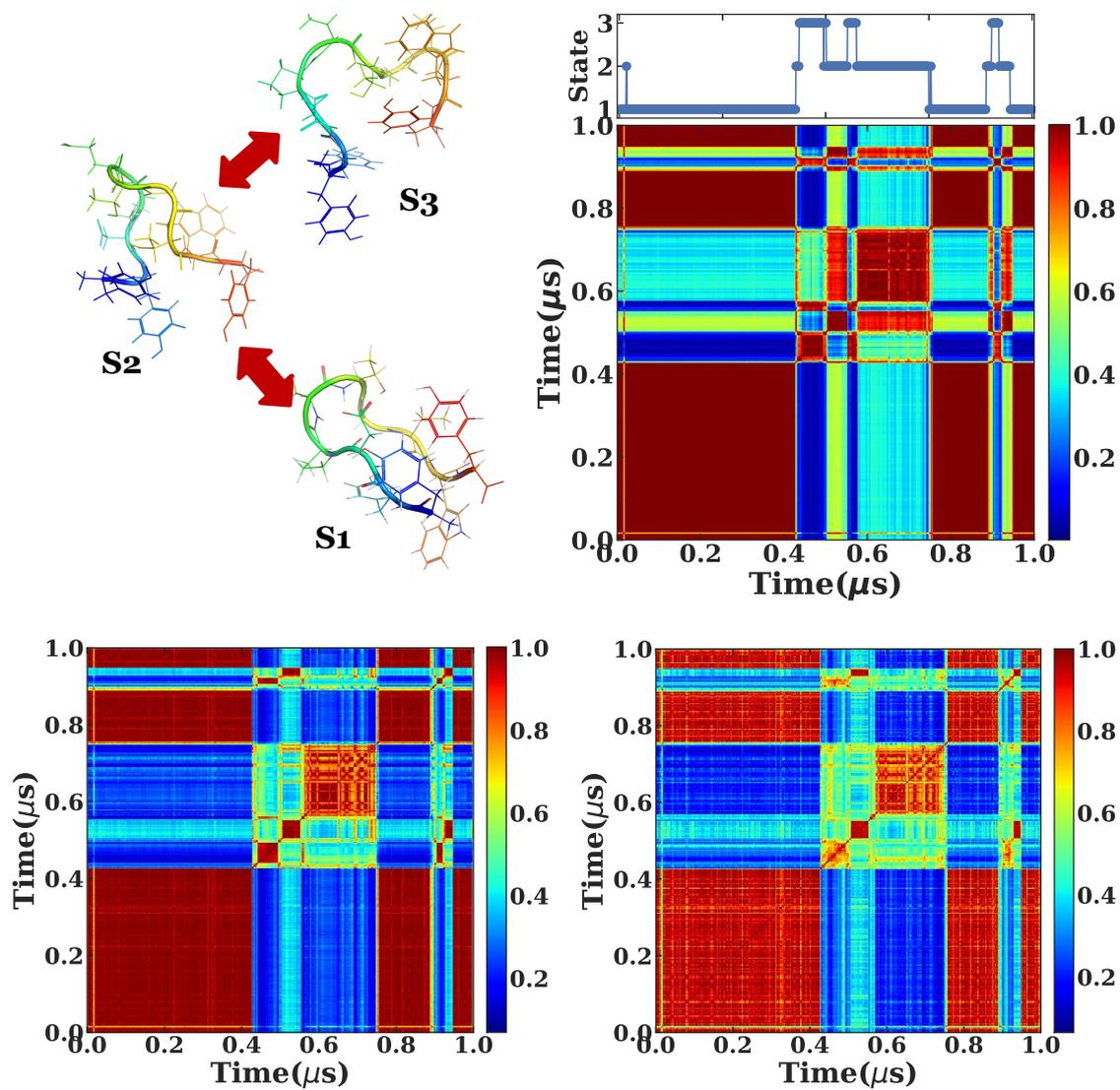

**Figure 3 State Transition and Similarity Matrices.**

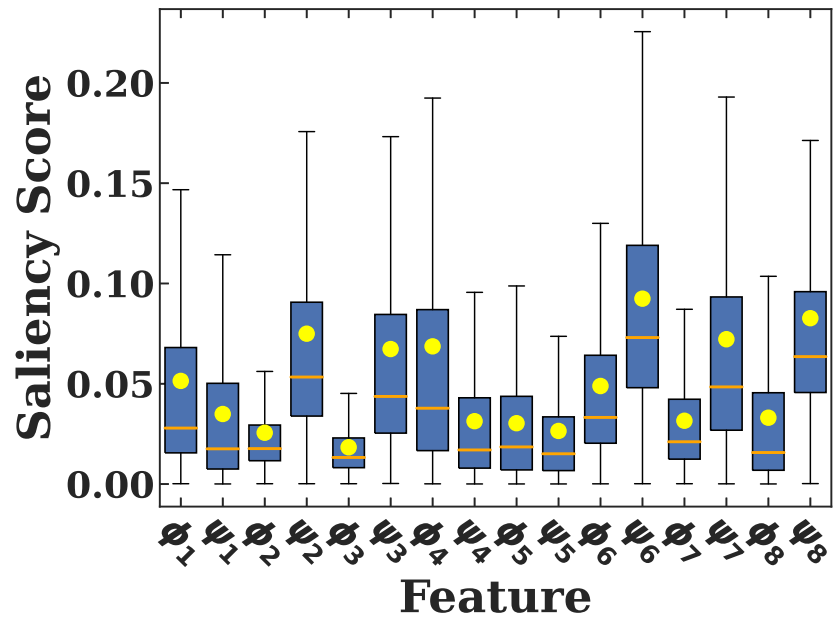

Figure 4 Saliency score of the neural network.